# LIPSS on thin metallic films: New insights from multiplicity of laser-excited electromagnetic modes and efficiency of metal oxidation

Alexander V. Dostovalov,[1,2] Thibault J.-Y. Derrien,[3,4] Sergei A. Lizunov,[3,5] F Přeučil,[3,6] Konstantin A. Okotrub,[1] Tomáš Mocek,[3] Victor P. Korolkov,[1,2] Sergei A. Babin[1,2] and Nadezhda M. Bulgakova[3,5,*]

[1] Institute of Automation and Electrometry of the SB RAS, 1 Acad. Koptyug Ave., 630090 Novosibirsk, Russia

[2] Novosibirsk State University, 2 Pirogova St., 630090 Novosibirsk, Russia

[3] HiLASE Centre, Institute of Physics, Czech Academy of Sciences, Za Radnicí 828/5, 25241 Dolní Břežany, Czech Republic

[4] Max Planck Institute for Structure and Dynamics of Matter (MPSD), Luruper Chaussee 149, 22761 Hamburg, Germany

[5] S. S. Kutateladze Institute of Thermophysics SB RAS, Lavrentyev Ave. 1, Novosibirsk 630090, Russia

[6] Institute of Particle and Nuclear Physics, Faculty of Mathematics and Physics, Charles University, V Holešovičkách 2, 180 00 Praha 8, Czech Republic

* corresponding author: bulgakova@fzu.cz

**Abstract**. Thin Cr films 28-nm thick deposited on glass substrates were processed by scanning low-intensity femtosecond laser pulses with energy well below single-pulse damage threshold. Two types of laser-induced periodic surface structures (LIPSS) were produced, depending on the scanning velocity, (1) parallel to laser light polarization with periodicity somewhat smaller than laser wavelength and (2) perpendicular to polarization with spatial period much smaller than wavelength. All structures are formed as protrusions above the initial film surface and exhibit a high degree of oxidation. To explain formation of the LIPSS and their conversion from one to another type, a rigorous numerical approach for modeling surface electromagnetic waves in thin-film geometry has been developed, which takes into account the change of optical properties of material due to laser-induced oxidation and porosity. The approach addresses the multiplicity of electromagnetic modes allowed for thin films. It has been found that the low spatial frequency LIPSS parallel to laser polarization, which are formed at low scanning velocities, are well described by the Sipe theory for surfaces of low roughness. The SEW mode responsible for high spatial frequency LIPSS formation at high scanning velocities has been identified. The mechanisms of optical feedback and transformation between types of LIPSS with scanning velocity have been proposed.

1. **Introduction**

Laser-induced periodic surface structures (LIPSS) attract much attention since their discovery on semiconductor surfaces in 1965 by Birnbaum [1]. They represent a periodic surface relief linked with polarization of laser light. During more than 50 years of investigations, these structures have found several niches of applications (modification of surface wetting properties [2], tribology [3], surface coloring and marking [4], enhancing of performance of implants [5], growth of living cells [6]). However, there are still considerable debates on the mechanisms of LIPSS formation [7-13]. The reasons for disagreements on the origin of periodical surface structuring are due to the fact that, by the date, a unified theory does not exist, which would quantitatively describe LIPSS periodicity, orientation relative to laser polarization, and quality for a great variety of materials under various laser-irradiation conditions.

Two main types of the LIPSS have been identified, low spatial frequency LIPSS (LSFL) and high spatial frequency LIPSS (HSFL) [9]. Both of these types can be formed either parallel or perpendicular to laser beam polarization [9] and can be conventionally divided to subfamilies depending on periodicity and depth-to-period aspect ratio [14]. Furthermore, for metals and semiconductors, the LIPSS are characterized as metal-oxide [10] or thermochemical [15-17] and "ablative" [12,18,19], the latter show a very low oxidation degree. Puerto et al. [20] have shown recently that highly periodic LIPSS on silicon can represent alternation of crystalline and amorphous phases. The LIPSS produced on transparent glasses and crystals have demonstrated an analogy in the formation mechanism with volumetric nanogratings [21]. Such a wide variety of periodic surface structures formed on surfaces of different materials, though linked with light polarization, implies differences in the formation mechanisms and in the processes involved in the formation that hinders understanding of underlying physics and, hence, their controllable fabrication.

The LSFL formation is reasonably described based on the interference of the incident laser wave and a surface electromagnetic wave (SEW) generated by scattering of laser light on the surface roughness relief, as established by Sipe et al. for metals [7] and generalized to other materials by Bonse et al. [22]. As a result of interaction between the incident laser beam and the SEW, a periodic pattern of the absorbed laser energy is created along the irradiated surface [7] which can further evolve into a periodic surface relief via various mechanisms, depending on the level of material heating. It was numerically proven [23] that the temperature modulation resulted from the periodic absorption pattern can survive a relatively long time to support periodic melting, ablation, and melt instability [12]. Other explanations of surface relief evolution into the LIPSS include instability-driven self-organization [24], in particular involving accumulation of surface electric charge [25]. It was demonstrated [19] that the regularity of ablative LSFL on metals can be well described quantitatively in terms of the SEW decay length along the surface. However, the origin of HSFL is not described by the Sipe theory [14] while some progress was achieved via numerical simulations [26]. The role of oxidation in LIPSS formation and their periodicity is also not yet understood.

In this paper, we address the problem of LIPSS formation on thin metallic films deposited on a transparent glass. The LIPSS were produced on 28-nm Cr films by scanning the film surface with femtosecond laser beam at laser fluences below the Cr single-pulse damage threshold. They can be classified as metal-oxide or thermochemical LIPSS [10,15-17] as the final structures contain considerable fraction of oxide. The central goal of this study is to uncover an intriguing effect, the transition from the LSFL to the HSFL, which are formed respectively parallel and perpendicular to laser light polarization, with increasing velocity of scanning laser beam. For analyzing the experimental results, an advanced model has been developed based on the SEW dispersion relation for thin films deposited on a dielectric substrate. The model accounts for the interaction between the SEW excited at two film interfaces, with air and glass substrate. The analysis reveals that the periodicity and orientation of the LSFL are reasonably described by the Sipe theory when accounting for the actual surface roughness. An optical feedback mechanism for pulse-by-pulse oxidation has been proposed to explain formation of high oxidation ridges with non- or slightly-oxidized metal regions between ridges. An advanced thin-film model with taking into account the oxidation of metal has enabled to identify the SEW mode which can be responsible for HSFL formation on the top of the LSFL. It has been demonstrated that film porosity plays a significant role in the HSFL periodicity. The mechanism of transition from the LSFL formed at low scanning velocities to the HSFL, which cover the total laser scanning track at higher scanning velocities, has also been proposed.

## 2. Experimental arrangement and results

The experiments on the LIPSS formation were carried out on the setup described in detail in [15]. Shortly, femtosecond laser pulses (PHAROS 6 W laser from Light Conversion, Inc.) with a central wavelength of 1026 nm, duration of 232 fs, and repetition rate of $f$ = 200 kHz were focused by a lens with the focal distance of 35 mm providing the beam diameter $d \approx$ 15 µm on the sample surface. The pulse energy $E_p$ was in the range 100-110 nJ. The samples were translated perpendicularly to the laser beam with the translation velocity, which was varied in a wide range from 1 µm/s to 300 µm/s. Metal films were deposited on BK7 borosilicate glass substrates by the magnetron sputtering technique which provides controlled film thickness $h$ of 28 nm. Scanning electron microscopy (SEM) Hitachi TM3000 was used to inspect nanostructures formed on the sample surface. Raman spectra of the laser processed surfaces were obtained using an imaging spectrometer (SP2500i, Princeton Instruments) with spectral resolution 2.5 cm$^{-1}$. The wavelength and power of pump radiation was 532.1 nm and 4 mW respectively (solid-state laser Excelsior, Spectra Physics).

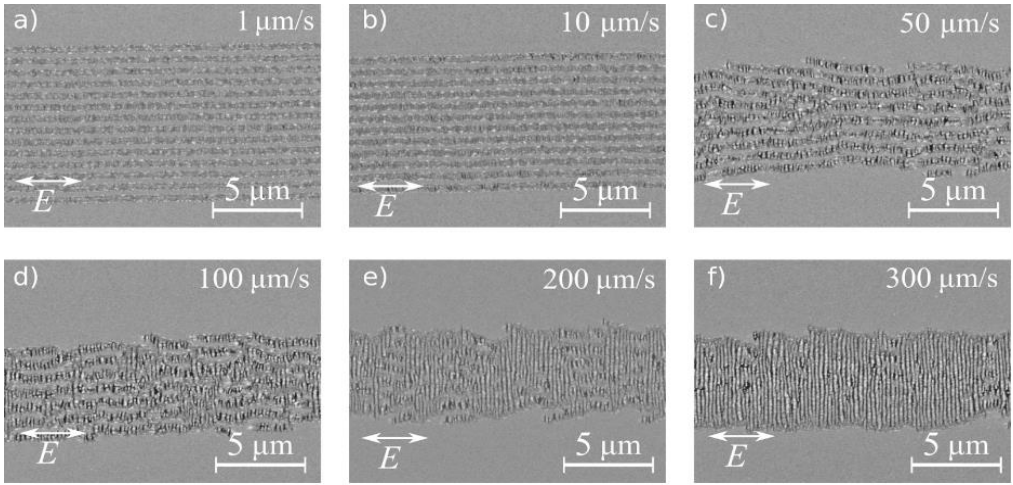

**Fig. 1**. SEM images of laser-nanostructured chromium films with $h$ = 28 nm at different scanning velocities: 1 µm/s (a), 10 µm/s (b), 50 µm/s (c), 100 µm/s (d), 200 µm/s (e), 300 µm/s (f). Laser pulse energy was 110 nJ. Scanning direction was along the light polarization shown by the double-sided arrows.

The LIPSS produced on 28-nm Cr films at different scanning velocities $v_s$ are presented in Fig. 1. The number of pulses coupled to the same area of the film, which were evaluated as N = $fd/v_s$, were $3\times10^6$, $3\times10^5$, $6\times10^4$, $3\times10^4$, $1.5\times10^4$, and $10^4$ for scanning velocities $v_s$ of 1, 10, 50, 100, 200, and 300 µm/s respectively. Highly regular periodic structures oriented along the laser polarization direction with the period of ~680 nm are formed at low $v_s$ of 1 µm/s - 10 µm/s (Fig. 1(a)-(b)). They can be classified as low spatial frequency LIPSS (LSFL) [9,14]. It should be underlined that the LSFL produced in this work grow up above the initial surface of the film [17], contrary to those formed at the regimes of strong ablation when the peaks of the LIPSS ridges occur below the initial surface [27]. In our cases, even inter-ridge regions remain at or above the initial surface level (see Fig. 3 in [17]).

Increasing $v_s$ to 50 µm/s yields in reducing the regularity of the LSFL with an increase in their period (Fig. 1(c)) to approximately 820 nm. Under these conditions, on the top of LSFL, the fine periodic structures with the direction perpendicular to laser light polarization becomes pronounced. Their period is in the range of ~200-250 nm and, hence, they can be referred as the high spatial frequency LIPSS or HSFL [9,14]. Increasing $v_s$ to 100-200 µm/s results in merging the adjacent ridges of the LSFL (Fig. 1(d)-(e)). Upon merging, the regularity of the HSFL, which are now covering the whole laser irradiation track, increases as clearly seen for $v_s$ = 300 µm/s (Fig. 1(f)). It should be noted that the similar tendencies are observed on the thicker films [17]. Shortly, at low scanning velocities, the LSFL parallel to laser polarization are formed on the Cr films with the thickness in the range 28-350 nm. The height of the LSFL ridges can be comparable or even larger than the film thickness while between the LSFL ridges the film remains at almost its initial level for the films up to 125 nm thick [17]. Hence, here we focus on only one film thickness, 28 nm, anticipating that our theoretical analysis presented below is valid for thicker films.

The protrusion of the LSFL well above the initial surface level indicates strong oxidation of the irradiated metal in the ridge sites. Indeed, for low scanning velocities, the transmission measurements demonstrate a highly regular striped pattern of transmitted light with high transmission at the ridge areas and with suppressed transmission between the ridges (Fig. 2(a), 1 µm/s) where oxidation should be small. The corresponding profile across the scanning track (Fig. 2(d)) demonstrates that inter-ridge sites are also somewhat protruded above the initial surface indicating some oxidation. With increasing scanning velocity, transmission pattern becomes irregular (Fig. 2(b), 70 µm/s) and, at even higher velocities, the LSFL stripes are hardly distinguishable (Fig. 2(c), 250 µm/s). The profile across the scanning track becomes as a whole well protruded above the initial surface with random spikes (Fig. 2(e)).

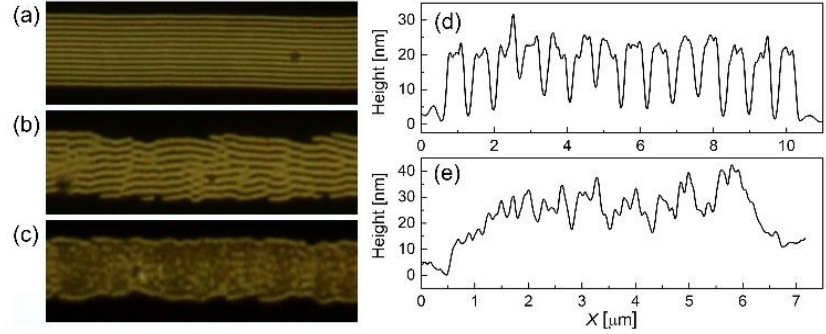

**Fig. 2**. White light transmission images through the laser-processed Cr films 28 nm thick at different scanning velocities, 1 μm/s (a), 70 μm/s (b), 250 μm/s (c). Profiles of the structures across the scanning tracks (a) and (c) are shown in (d) and (e) respectively.

Raman measurements show the presence of two oxides, $CrO_2$ and $Cr_2O_3$, in all irradiation regimes (Fig. 3). Interestingly, the formation of half-metallic chromium(IV) oxide ($CrO_2$) dominates at the present irradiation conditions. It is known that $CrO_2$ converts naturally into more stable oxide $Cr_2O_3$. Being exposed to oxygen or air under normal conditions, $CrO_2$ is usually covered by 2-3 nm layer of $Cr_2O_3$ [28] which preserves chromium(IV) from further conversion. Such $Cr_2O_3$ layer provides high corrosion resistance to chromium whose efficient oxidation starts only at temperatures of 400-450°C and higher. At low scanning velocities, the ratio $Cr_2O_3/CrO_2$ stays in the range of 0.3-0.4 and is increasing at higher scanning velocities (Fig. 3(b)). These data will be used below for the theoretical analysis of LIPSS formation.

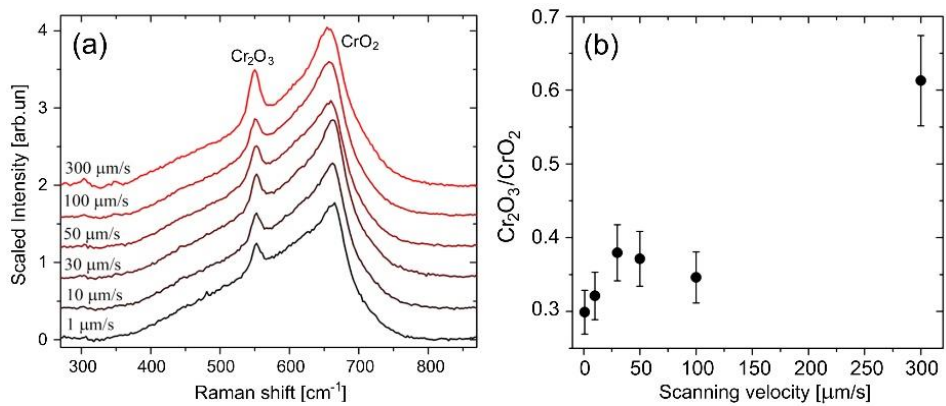

**Fig. 3**. (a) Raman spectra of the LIPSS at different scanning velocities demonstrating the presence of two oxides, $Cr_2O_3$ and $CrO_2$. (b) Composition ratio of the two oxides.

### 3. Theoretical analysis

3.1 *Low spatial frequency LIPSS*

As shown in the previous Section, in all irradiation regimes studied here, the Cr films are strongly oxidized by analogy with observations in [10]. At low scanning velocities, the formed LSFL (Figs. 1-2) are parallel to laser light polarization that can be attributed to dipole-like scattering of the laser beam on surface roughness (surface scattering centers) as proven theoretically in [10] (see also Supplementary information for [10]). Here we apply the Sipe theory, to explore the periodicity and orientation of LIPSS produced in our experiments, by calculating the efficacy factor $\eta$. This factor characterizes the efficacy, at which laser light coupling with the surface roughness results in inhomogeneous (periodic) laser energy absorption $A(\vec{k})$ with $\vec{k} = (k_x; k_y)$ to be the vector of the induced periodic structure [7]: $A(\vec{k}) \propto \eta(\vec{k}, \vec{k_i}) b(\vec{k})$. Here $\vec{k_i}$ is the component of the laser wave vector parallel to the surface representing a "dipole sheet" composed of roughness features and $b(\vec{k})$ is the Fourier component of the roughness. The $b(\vec{k})$ value is described by two parameters, $F$ and $s$, which are respectively referred to as the filling and shape factors [29]. The $\eta$ values can be derived analytically or numerically under the assumptions that the roughness features are much smaller as compared to the laser wavelength [7,22,29].

Figure 4 presents the efficacy factor map $\eta(k_x, k_y)$, calculated for 1026 nm wavelength based on the optical properties for pure chromium [30]. The filling factor $F$ was taken to be 0.1, accounting for a low roughness (≤5 nm) of the films deposited by the magnetron sputtering technique. The shape factor $s$ was chosen to be 0.4 that corresponds to spherically shaped islands [22,29]. Important is that, in the case of low roughness, an extended maximum in the efficacy factor map is observed along $k_y$ direction (Fig. 4(b)), which is also marked by blue rectangle in Fig. 4(a)). This feature indicates dominance of LIPSS formation in the direction parallel to light polarization, in line with [10]. The $\eta$ maximum along $k_y$ direction ($k_x = 0$), which is located at $k_y = 1.278$, corresponds to the LSFL periodicity of 802 nm, in reasonable agreement with the present observations of the LSFL with the period in the range of 680-910 nm, depending on the irradiation conditions. This brings us to an important conclusion that the LSFL formation can be initiated on yet purely metallic surface via periodic absorption of the laser light by metallic film. However, imprinting of the periodic absorption profile onto the film surface proceeds via oxidation along the absorption lines of the enhanced temperature (see discussion below).

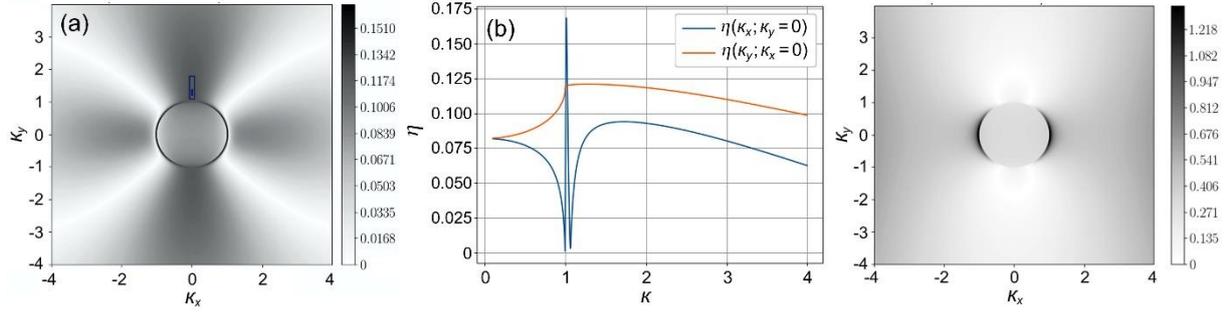

**Fig. 4**. (a) The efficacy factor map of LIPSS formation for the interface between air and chromium surface calculated for 1026 nm wavelength based on the optical properties from [30]. Blue rectangle with the dot inside marks the region of $k_y$ with maximum $\eta$, which corresponds to the most probable efficiency for the LSFL formation. (b) Efficacy factor along $k_x$ and $k_y$ corresponding to (a). (c) The same as (a) for filling factor of 0.9 (rough surface). For all calculations, the shape factor was taken equal to 0.4 that corresponds to spherical scattering centers [22,29].

A sharp peak of the $\eta$ value along $k_x$ direction at $k_x = 1$ (Fig. 4(b)), though having app. 35% larger amplitude than the $\eta$ maximum along $k_y$, is however very narrow, anticipating a low probability for realization of the LSFL perpendicular to light polarization (only a very narrow range of wave-numbers can enable the coupling of the far field laser light with the selvedge near-field). It is worth noting that, for larger filling factors, above 0.5 (rough surface), the calculated efficacy factor indicates dominance of the LSFL perpendicular to light polarization (Fig. 4(c)). We underline here that the HSFL are not predicted by the Sipe theory and their formation will be analyzed below.

3.2 *Imprinting LSFL via local laser-enhanced oxidation*

If the LSFL parallel to laser polarization are initiated by light scattering and by the formation of a corresponding periodic absorption profile on a *pure metallic* surface, the question arises on how they are imprinted onto the film surface, in particular via involving oxidation. Important is also to elucidate the effect of increasing the scanning velocity, which transforms the LSFL to the HSFL (Figs. 1-2). As the oxidation processes are strongly temperature-dependent, it is necessary to estimate the level of film heating by pulsed laser irradiation as well as the effect of temperature modulation in the irradiated film resulted from periodic light absorption [23].

Estimations indicate that in our case the LIPSS formation should proceed well below the ablation threshold and, under certain conditions, it can take place below the melting threshold. For metal films with thickness of 28 nm studied here, 28 nm, assuming that the total absorbed laser energy is confined within the film due to a low thermal conductivity of glass substrate, the melting temperature ($T_m$ = 2180 K) is not obligatory reached. Indeed, the temperature increase as a result of the pulse action on a thin film of the thickness $h$ can be expressed as

$$\Delta T = (1 - R)F_0(1 - e^{-\alpha h})/c_l h \qquad (1)$$

where $F_0$ is laser fluence, $R$ and $\alpha$ are the reflection and absorption coefficients, $c_l$ is the heat capacity of chromium. Here it is assumed that the temperature is quasi-uniform across the nanosized film due to the high heat conduction of the electronic subsystem that was justified by numerical modeling [31]. As the heat capacity of chromium is noticeably growing upon heating [32], by using an iterative procedure, we obtain the maximum temperature of 28-nm film in the range of ~1900-2000 K under the present irradiation conditions. Note that the maximum temperature drops with the film thickness (Eq. (1)). Thus, for thicker films, the maximum temperature reached in the film as the result of single pulse action will be even lower. Important is that a noticeable heat accumulation in the film cannot be expected at the repetition rate used in our study. Indeed, a typical heat conductivity of glass materials provides the propagation rate of the temperature front of the order of 1 μm per 1 μs [33]. With the pulse repetition rate of 200 kHz (separation time between pulses of 5 μs), the heat absorbed by a nanosized film must considerably dissipate into glass substrate between successive pulses. However, for thinnest films studied here, by accounting for periodic modulation of absorption profile along the surface, periodic melting in the absorption peaks cannot be excluded [23]. We note that, for an astigmatic Gaussian beam [34], heat accumulation is more pronounced due to different geometry of heat dissipation compared to the present case.

It is well known that pure chromium is distinguished by very high corrosion resistance under normal conditions due to formation of thin oxide film on the surface. However, already starting from temperatures of 400-450 °C, chromium is efficiently oxidized in air. The oxidation process proceeds due to diffusion of metal cations through the growing oxide layer to the sample surface while the oxygen anions diffuse in the opposite direction [35,36]. This process is enhanced at high temperatures and especially in the molten state [37]. Thus, in air under atmospheric pressure, the parabolic rate constant can exceed $2\times10^{-12}$ m$^2$/s at 1600 K and can be extrapolated to ≥$10^{-10}$ m²/s at temperatures of 1900-2000 K [38]. At such oxidation rates, it can be estimated that, under our irradiation conditions, chromium can be oxidized to the depth of several hundreds of nanometers. Indeed, assuming that the film preserves the temperature of ~1500-1600 K up to ~10 ns after the laser pulse that is reasonable in view of relatively low conductivity of the underlying grass substrate, for $10^6$ pulses per spot size (total time of the film in the hot state of 0.01 s) we estimate the oxidation depth >100 nm. Note that the temperature can be transiently even higher, especially in the absorption peaks, and the oxidation process continues also during slow cooling between sequential pulses that increases the oxide thickness. This explains why thin films considered in this study are oxidized throughout as seen in Fig. 2. It is necessary to stress that laser irradiation is known to efficiently enhance metal oxidation, which is caused by electron excitation and formation of temperature gradients [39]. Important is that laser-induced oxidation of chromium proceeds with formation of an ultra-porous layer [40], which favors the diffusion of metal cations and oxygen anions. Here we emphasize an additional mechanism of the positive feedback, which must considerably facilitate formation of the periodic oxide ridges once the oxidation process has been initiated by periodic absorption of the laser light.

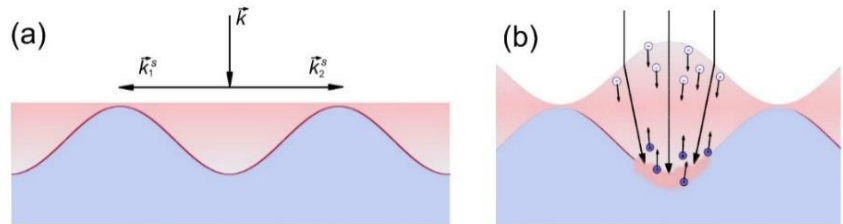

**Fig. 5**. Schematics of the feedback mechanism of LSFL formation via oxidation. (a) Laser light scattering by the surface roughness leads to the formation of the periodic absorption profile with the modulated temperature along the surface (hot regions are marked by pink). Here $\vec{k}$ is the wave vector of incident laser light; $\vec{k}_1^s, \vec{k}_2^s$ are the wave vectors of the SEW. (b) In the hot lines of absorption maxima, oxidation is more efficient compared to absorption minima (blue color represents cold areas). Due to consumption of oxygen from air for the formation of oxide, the hot lines protrude above the initial surface level. The metal cations (blue circles) diffuse through the growing oxide layer to the sample surface while the oxygen anions (white circles) diffuse in the opposite direction, thus enhancing oxidation.

The feedback mechanism is schematically illustrated in Fig. 5. Incident laser light is scattered by the surface roughness that leads to a periodic absorption profile with modulated temperature along the surface (Fig. 5(a)). Note that, as analyzed above, the absorption lines are oriented along laser polarization (Fig. 4). The modulation can reach 15-20% at similar irradiation parameters [23,41]. In the hot lines of the absorption maxima, oxidation is much more efficient compared to the lines of the absorption minima due to a non-linear nature of the oxidation process [38]. The metal cations diffuse through the growing oxide layer to the sample surface while the oxygen anions diffuse in the opposite direction [35,36], thus enhancing oxidation (Fig. 5(b)). Due to consumption of oxygen from air for oxide formation, the hot lines protrude above the initial film surface while the porous structure of the oxide protrusions favors oxidation in these specific protrusion regions. Additionally, protruded oxide lines may serve as cylindrical nano/microlenses, which focus next laser pulses toward oxide bottom, thus enhancing temperature modulation along the film and facilitating periodic oxidation.

We note that the considerable modulation of the temperature on the surface can be preserved up to several hundreds of picoseconds [23] while oxide cylindrical nano/microlenses can further prolong this time and enhance modulation amplitude that calls for further studies.

3.3 *Origin of the HSFL*

As reported above, with increasing scanning velocity, the secondary periodic structure with high spatial frequency appears on the top of the oxide ridges with orientation perpendicular to laser light polarization. Furthermore, this structure starts to dominate at highest scanning velocities and completely masks the LSFL (Fig. 1). To understand the origin of this additional structure, which can be referred as the HSFL, below a rigorous analysis is performed based on the SEW dispersion relation for a thin metallic film deposited on a glass substrate. We solve an implicit equation based on the dielectric permittivities $\varepsilon_1$, $\varepsilon_2$, and $\varepsilon_3$, associated with the propagation wave-vectors $k_1$, $k_2$, and $k_3$, respectively. The propagation vectors are defined by the light momentum conservation expressed as

$$k_j^2 = \beta^2 - \varepsilon_j \left(\frac{\omega}{c}\right)^2 \qquad (2)$$

where $j = 1$, 2 or 3 related to the film material (chromium in our case as well as a mixture of chromium with chromium oxide in the case of oxidation), air, and the substrate respectively; $c$ is the light velocity in vacuum/air; $\omega$ is the laser frequency. Note that here the problem is being solved for the more general case than in [42] where the solution was limited to a purely metallic film. The periods of possible modes of the surface electromagnetic waves resulting from the interference of the laser wave with the SEW is provided by the relation $\Lambda = 2\pi/Re(\beta)$ where $\beta$ is the complex wave number associated with the SEW modes. The periods can be calculated by numerically solving Eq. (2) with the following equation [43]

$$e^{-2k_1 t} = \left(\frac{k_1}{\varepsilon_1} + \frac{k_2}{\varepsilon_2}\right)\left(\frac{k_1}{\varepsilon_1} + \frac{k_3}{\varepsilon_3}\right) / \left(\frac{k_1}{\varepsilon_1} - \frac{k_2}{\varepsilon_2}\right)\left(\frac{k_1}{\varepsilon_1} - \frac{k_3}{\varepsilon_3}\right) \quad (3)$$

Prior to solving Eq. (3) for media $j = 1,2,3$, we express $k_j$ using

$$k_j = s_j\sqrt{\beta^2 - \varepsilon_j\left(\frac{\omega}{c}\right)^2} \quad (4)$$

where $s_j$ is the sign of the square root equal to + or -. As we have three media, $2^3 = 8$ choices can be made for the definition of complex-valued SEW wave-vectors $k_j$ [43]. Note that the change the sign $s_1$ in Eq. (4) keeps Eq. (3) invariant. Thus, the number of cases is reduced to 4, which are characterized by the choice of signs made for pairs $(k_2, k_3)$ which we classify as (+,+), (+,-), (-,+), and (-,-) and associate with the propagative nature of the SEW modes on metallic films [44,45]. In the convention we use here (an external environment is medium 2, a thin metallic/oxide film is medium 1, and a substrate is medium 3, see Fig. 6), the branch $(k_2, k_3) =$ (+,+) is associated with the SEW which are not radiating to the far field, i.e., the waves bound to the material surface. The branches (-,+) and (+,-) are associated with "leaky modes" as they radiate energy to the far field, above and below the film respectively. For the branch (-,-), the fields grow exponentially away from interfaces and, thus, this branch has a leaky character. The modes of this branch were first analyzed for metallic slabs based on Maxwell's equations in [45,46]. The mode (-,-) shown in Fig. 6 has been classified by Norrman et al. [45] as a high order mode. Such leaky waves were earlier considered as unphysical because of unlimited growing of their amplitude away from the metal film boundaries. However, as noticed in [44,45], they can be meaningful in the transient sense over a limited region of space in the nearest vicinity of the film interfaces. All various solutions of Eq. (3) can be regarded as resonant modes in the geometry of a thin film interfacing with dielectric media with different dielectric permittivities [45]. Below we show these solutions for the metal film which experiences oxidation and demonstrate how the modes of the electromagnetic waves, which are excited on two film interfaces and mutually interacting, evolve with the degree of oxidation.

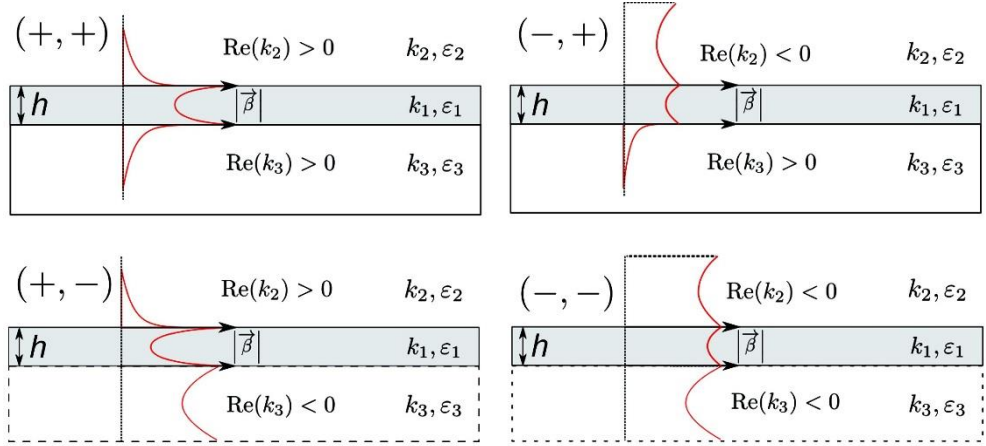

**Fig. 6**. Physical meaning of the branches of the SEW modes. Figure was inspired by [44,45]. For explanations, see the text.

The calculations of Eq. (3) were performed with varying the oxidation degree of the metal film from 0 to 100%. We consider the fixed ratio of $Cr_2O_3$ fraction to that of $CrO_2$ to be equal to 0.35, which is within the typical oxidation range for the moderate scanning velocities (Fig. 3). The dielectric function of chromium as a function of the oxidation level was calculated within the concept of the Lorentz–Lorenz effective medium [47] as follows. For $CrO_2$ as uniaxial oxide, we have adapted the data from [48] reported for the directions both perpendicular and parallel to c-axis. For 1026 nm wavelength, they were evaluated as $0.58 + 6.48i$ and $1.36 + 9.01i$ respectively. Then, the dielectric function of $CrO_2$ was calculated by the Lorentz–Lorenz approach for random orientation of crystallites (2:1 ratio between ordinary and extraordinary dielectric functions). The obtained dielectric function was used for calculating the dielectric function of the $Cr_2O_3/CrO_2$ mixture of 0.35 ratio, using $\varepsilon = 3.827 + 0.048i$ adapted for $Cr_2O_3$ at 1026 nm from [49]. Finally, simulations of the dielectric function of chromium as a function of oxide fraction in the film were performed whose results are presented in Fig. 7 by solid lines. Additionally, it was assumed that laser-produced chromium oxide is usually porous as was found in [40] that can influence the LIPSS periodicity [16]. Dashed lines in Fig. 7 illustrate how the dielectric function of the oxidized chromium changes when 20% of porosity ($\varepsilon = 1$) is added to the same oxide mixture, $Cr_2O_3/CrO_2$, of 0.35 ratio.

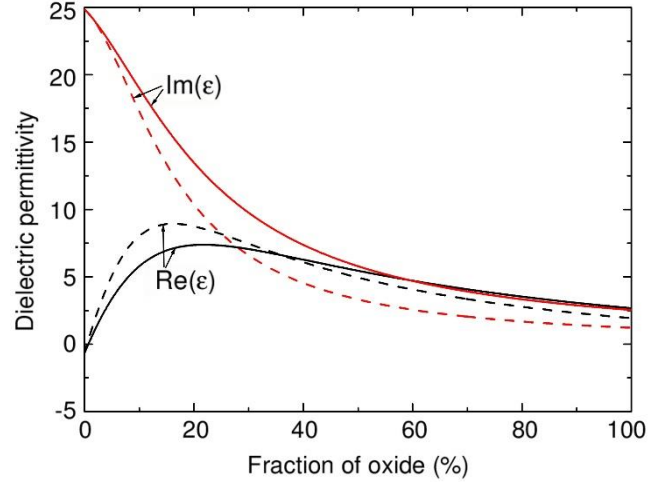

**Fig. 7**. The dielectric function at the wavelength of 1026 nm for the mixture of chromium with chromium oxides. Solid lines correspond to the fixed compositional ratio of two oxides, $Cr_2O_3/CrO_2$, of 0.35 as revealed by Raman analysis of the structures produced at low scanning velocities (Fig. 3(b)). For calculations of the dashed lines, it was assumed that the mixture of two oxides, $Cr_2O_3/CrO_2$, with the compositional ratio of 0.35 is porous with 20% of volumetric porosity. The curves are obtained within the concept of the Lorentz–Lorenz effective medium [47].

Equation (3) was solved by using two methods, the modified Powell method and the Levenberg-Marquardt method [42]. The solution for the metal film having two interfaces with dielectric media corresponds to all possible combinations of solutions for each metal interface as illustrated in Fig. 6. The simulations were performed for a 28-nm Cr film on a glass substrate ($\varepsilon = \varepsilon_3 = n^2$ with refractive index $n$ of 1.45) for the case of the scattered wave vectors $k_j$ ($j =$ 1,2,3) parallel to laser light polarization in order to reveal the origin of the HSFL whose ridges are perpendicular to polarization. As a result, the multiple roots are obtained for each of the cases presented in Fig. 6. Among the possible wave-numbers $\beta$, we consider all solutions provided by the numerical solver, which can lead to several SEW modes, simultaneously possible for a given material. The results of simulations are presented in Fig. 8 in the form of periods of possible modes $\Lambda = 2\pi/Re(\beta)$ as a function of oxidation percentage for the oxide ratio $Cr_2O_3/CrO_2$ of 0.35. Note that namely the period of a dominating mode should be considered as the possible LIPSS periodicity. To illustrate the effect of change in the oxide ratio, the modes of the (-,-) branch are added for the case of chromium oxidation with $Cr_2O_3$ as the only oxide (Fig. 8, left). The simulations were repeated for the assumption of 20% porosity of the formed oxide with the results presented in Fig. 8, right.

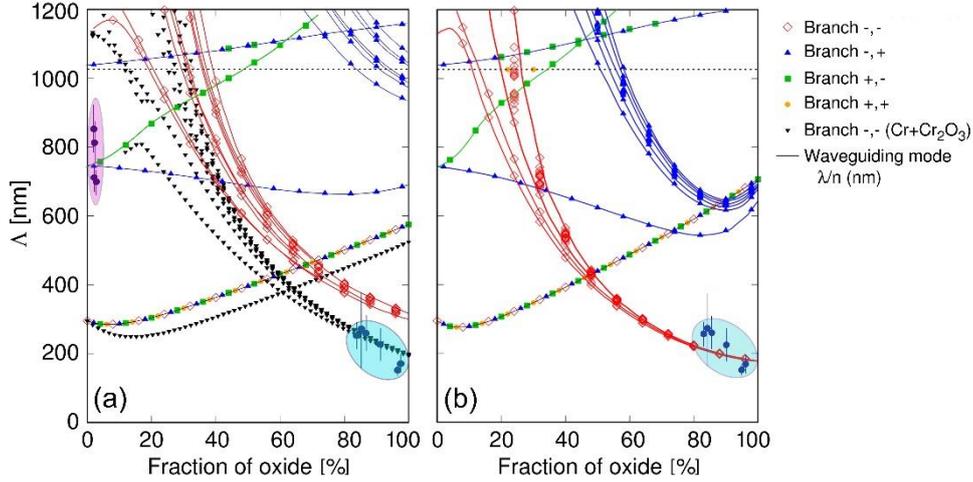

**Fig. 8**. Possible periodicities of LIPSS, which can be potentially formed due to the modes of the scattered SEW, as obtained by solving Eq. (3) for 28-nm chromium film on glass. Colored dots correspond to different kinds of modes obtained for chromium oxidized with the mixture of two oxides $Cr_2O_3$ and $CrO_2$ with the compositional ratio of 0.35, without and with 20% porosity on the left and on the right respectively. Black triangles pointing downwards are given for oxidation chromium only into the form of $Cr_2O_3$, showing how the periodicity corresponding to the leaky (-,-) mode depends on material composition. The waveguiding mode $\Lambda = \lambda/n$ is also shown. Right graph demonstrates a strong effect of porosity on the structure periodicity. Experimental points are shown by dots with error bars: the HSFL data are outlined by blue oval while the LSFL ones are shown inside pink oval.

By comparing the experimental results on the HSFL periodicity with the possible modes obtained via solving Eq. (3), we have found that they are reasonably well fitting the branch (-,-) at high oxidation degrees. The experimental data for 28-nm films on the LSFL and HSFL periodicities are summarized in Table 1 and shown in Fig. 8 (outlined by pink and blue ovals respectively). Note that the points for the LSFL are added optionally for illustration while their orientation does not correspond to the simulation geometry of the scattered SEW (see Section 3.1). When positioning the HSFL experimental points to Fig. 8, it was assumed that the oxidation is high for all points and it is increasing with decreasing the scanning velocity (note that the content of the metallic phase of chromium was not measured). This is a reasonable assumption as from Fig. 2 it is evident that the film becomes essentially transparent after laser scanning. Thus, the simulations point out that the HSFL, which are formed on the top of the oxidized regions, are plausibly the product of the (-,-) electromagnetic mode. For relatively thin films, this mode is exponentially growing and can become strong in a region nearby the film interfaces [44,45], dominating over other possible excited SEW.

**Table 1**. Summary of the experimental results on LIPSS periodicities obtained for 28-nm chromium film at different scanning velocities.

| Scanning velocity (μm/s) | LSFL average period (nm) | LSFL period error (nm) | HSFL average period (nm) | HSFL period error (nm) |
|---|---|---|---|---|
| 1 | 696 | 78 | 170 | 64 |
| 10 | 704 | 71 | 159 | 38 |
| 50 | 816 | 139 | 217 | 101 |
| 100 | 858 | 140 | 244 | 110 |
| 200 | - | - | 249 | 238 |
| 300 | - | - | 238 | 72.5 |

We underline here that the HSFL experimental points are positioned in Fig. 8 arbitrarily, considering higher oxidation degree for lower scanning velocities. In this respect, the behavior of periodicity of the HSFL, which can be caused by the (-,-) mode of the scattered electromagnetic waves on an oxidized metal film, with the scanning velocity (oxidation degree) is reasonably described qualitatively. The model for the measured ratio of the contents of $Cr_2O_3$ and $CrO_2$ oxides (Fig. 3) predicts approximately 1.5 times larger period than observed experimentally (Fig. 8, left). However, when assuming that the formed oxide is porous [16,40], a very good agreement has been obtained between the calculated (-.-) mode periodicity and the experimental data (Fig. 8, right). Thus, the metal oxidation degree together with porosity of the formed oxide look to be the major governing factors of the HSFL formation.

Other factors, which can influence the HSFL periodicity, should also be mentioned. All simulations here were performed for optical properties of chromium and chromium oxides at normal conditions. However, upon pulsed laser heating material optical properties are transiently changing. It was shown experimentally that the HSFL periodicity on titanium were decreasing with laser fluence [49] and, hence, with the electron temperature achieved during laser irradiation [50]. As seen from Fig. 8 (low oxidation side) *for a pure or slightly oxidized metal film*, the branches of the SEW modes (-,+), (+,-) and (+,+) could be a responsible explanation for the HSFL formation if to assume that the periodicities of these modes are decreasing as a result of laser action. However, the phenomenon of multi-pulsed ultrashort laser irradiation of matter is an intricate phenomenon, which involves numerous processes and calls for further studies.

3.4 *On transition from LSFL to HSFL by increasing the scanning velocity*

The effect of transition from the LSFL whose ridges are parallel to laser light polarization to those decorated with the perpendicular-to-polarization HSFL on the top and finally to covering the whole laser-scanning track by the HSFL with increasing scanning velocity remains intriguing. The most plausible explanation can be in the laser energy dose coupling to the same irradiation area in the beginning of scanning. In particular, acceleration time required for the motorized scanning stage to achieve its stable velocity can play a role in initiation of structuring. The mechanism of this effect is shown schematically in Fig. 9. At low scanning velocities, the stage is located at the initial point for a longer time. The middle part of the Gaussian laser beam delivers a high irradiation dose to the metal film and induces a heavy oxidation both across and along the film (zone 1). However, in the periphery area of the irradiation spot, the conditions of oxidation are also achieved, though gentler than in the middle part of the spot, with formation of the LSFL (zone 2) as schematically shown in Fig. 5. Namely formation of a periodically structured zone 2 looks to be a key feature for duplicating of the structure to new areas upon scanning through the mechanism proposed by Öktem et al. [10]. This case is shown in the scheme of Fig. 9 (note that at $v_s = 1$ μm/s the effective number of pulses on the same sample area is $10^6$). The periodic oxidation ridges created in zone 2 confine the light of consequent laser pulses (Fig. 5) and thus lead to gradually copying themselves to the virgin film regions in the forefront of the scanning beam.

With increasing the scanning velocity, the overheated zone 1 with throughout oxidation preserves while the periodically structured zone 2 is considerably decreasing due to a strong decrease in the irradiation dose in the periphery attributable to the Gaussian beam profile. As a result, the ridges in zone 2 becomes too short and cannot provide accurate self-reproducing to new film areas (Fig. 9, 50 μm/s). At even higher scanning velocities (300 μm/s in Fig. 9), zone 2 becomes negligibly small that results in copying of only zone 1 to new areas. We underline that all three images in Fig. 9 are presented with the same spatial scale. The fact that the modification track becomes narrower with the scanning velocity counts in favor of the proposed mechanism. Note that the width of the modification track at the largest scanning velocity coincides with the diameter of the zones 1 observed at slower scanning (compare images in Fig. 8, right).

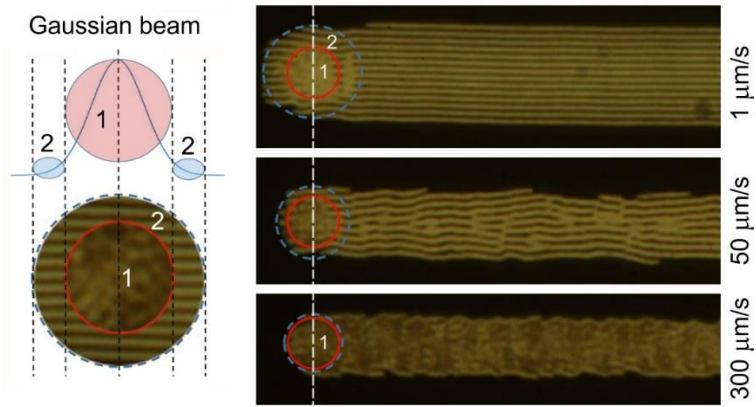

**Fig. 9**. Schematics of the mechanism responsible for transformation from the LSFL parallel to laser light polarization to the HSFL covering the whole scanned area, which is observed with increasing the scanning velocity. *Left*: Gaussian laser beam (top) can be arbitrarily divided into (1) the middle part (marked by pink circle) which overheats the metal film at the initial point of scanning and leads to almost uniform oxidation of the film and (2) beam periphery where gentle irradiation creates periodic oxidation ridges (marked by blue ovals). The structure of the starting point is sketchy shown in the left bottom. *Right*: At low scanning velocities, the peripheral zone 2 of the irradiation spot is covered by the oxidation ridges of the LSFL, which are ideally reproducing themselves to new areas in the leading edge of the scanning beam (top). With increasing scanning velocity, initial oxidation ridges in zone 2 become short as this zone does not receive enough radiation dose for their creation. As the result, copying the ridges to the new areas experiences fluctuations (middle). At high scanning velocities, zone 2 becomes negligibly small and only zone 1 can be reproduced to new areas (bottom). Note that the images on the right are given in the same spatial scale.

**Conclusion**

In this study, the periodic surface structures were produced on thin chromium films of a fixed thickness of 28 nm deposited on glass substrates by femtosecond laser beam scanning at the velocities from 1 to 300 μm/s. The energy of single pulses was below the film damage threshold and the effective number of pulses on the same irradiation area was in the range of $10^4$–$10^6$. All periodic structures demonstrate high level of metal oxidation that was supported by Raman measurements. At low scanning velocities, the metal-oxide LSFL are formed, whose ridges are parallel to laser light polarization and protrude above the initial film surface. With increasing scanning velocity, the HSFL perpendicular to light polarization emerge on the top of the LSFL ridges. At the highest scanning velocities used in this study, the LSFL ridges are merging into a continuous oxide region across the laser-irradiation track, which becomes finally covered by the HSFL.

To understand the mechanisms of the LSFL and HSFL, we have performed a rigorous theoretical analysis, which involves considerations of the actual experimental conditions and is based on the Sipe theory and simulations of the SEW dispersion relation for the thin film geometry. It was found that the periodicity of the LSFL parallel to light polarization is well described by the Sipe theory. This points out that they are initiated on the pure metallic film via while their imprinting onto the film proceeds through efficient periodic oxidation of metal in the regions of the light absorption maxima. The chemistry of metal oxidation has been discussed. The optical feedback mechanism for developing the periodic oxidation has been proposed.

As the HSFL are not described within known theories, the model for thin metallic films on dielectric substrate has been developed based on the dispersion relation, which includes laser-induced metal oxidation and oxide porosity. Simulations of the SEW dispersion relation have yielded in numerous modes of the surface electromagnetic waves, which can be imprinted on the metal-oxide films. It has been found that the leaky (-,-) mode of the SEW well describes the HSFL formation on the oxidized metal films. Finally, the mechanism responsible for the transition from the LSFL to the HSFL with increasing the scanning velocity has been proposed.

**Acknowledgments**

The research of A.V.D., V.P.K. and S.A.B. is financed by the state budget of Russian Federation (IA&E project №AAAA-A17-117062110026-3). The research of T.J.-Y.D., F.P., T.M. and N.M.B. is financed by the European Regional Development Fund and the state budget of the Czech Republic (project BIATRI, No. CZ.02.1.01/0.0/0.0/15_003/0000445; project HiLASE CoE, No. CZ.02.1.01/0.0/0.0/15_006/0000674; programme NPU I, project No. LO1602).